# *Highly tunable room-temperature plexcitons in monolayer WSe$_2$ /gap-plasmon nanocavities*

Thomas P. Darlington[1,†], Mahfujur Rahaman[2,†], Kevin W.C. Kwock[3], Emanuil Yanev[1], Xuehao Wu[4], Luke N. Holtzman[5], Madisen Holbrook[1], Gwangwoo Kim[2,6], Kyung Yeol Ma[7], Hyeon Suk Shin[7], Andrey Krayev[8], Matthew Strasbourg[9], Nicholas J. Borys[9], D. N. Basov[4], Katayun Barmak[5], James C. Hone[1], Abhay N. Pasupathy[4], Deep Jariwala[2], P. James Schuck[1*]

[1]Department of Mechanical Engineering, Columbia University, New York, NY 10027, USA
[2]Department of Electrical and Systems Engineering, University of Pennsylvania, Philadelphia PA 19104, USA
[3]Department of Electrical Engineering, Columbia University, New York, NY 10027, USA
[4]Department of Physics, Columbia University, New York, NY 10027, USA
[5]Department of Applied Physics and Applied Mathematics, Columbia University, New York, NY 10027, USA
[6]Department of Engineering Chemistry, Chungbuk National University, Chungbuk, 28644, Republic of Korea
[7]Department Chemistry, Ulsan National Institute of Science and Technology, Ulsan 44919, Republic of Korea
[8]Horiba Scientific, Novato, CA 94949, USA
[9]Department of Physics, Montana State University, Bozeman MT, 59717, USA
† These authors contributed equally to the work
* p.j.schuck@columbia.edu

**Abstract:**

**The advancement of quantum photonic technologies relies on the ability to precisely control the degrees of freedom of optically active states. Here, we realize real-time, room-temperature tunable strong plasmon-exciton coupling in 2D semiconductor monolayers enabled by a general approach that combines strain engineering plus force- and voltage-adjustable plasmonic nanocavities. We show that the exciton energy and nanocavity plasmon resonance can be controllably toggled in concert by applying pressure with a plasmonic nanoprobe, allowing *in operando* control of detuning and coupling strength, with observed Rabi splittings >100 meV. Leveraging correlated force spectroscopy, nano-photoluminescence (nano-PL) and nano-Raman measurements, augmented with electromagnetic simulations, we identify distinct polariton bands and dark polariton states, and map their evolution as a function of nanogap and strain tuning. Uniquely, the system allows for manipulation of coupling strength over a range of cavity parameters without dramatically altering the detuning. Further, we establish that the tunable strong coupling is robust under multiple pressing cycles and repeated experiments over multiple nanobubbles. Finally, we show that the nanogap size can be directly modulated via an applied DC voltage between the substrate and plasmonic tip, highlighting the inherent nature of the concept as a plexcitonic nano-electro-mechanical system (NEMS). Our work demonstrates the potential to precisely control and tailor plexciton states localized in monolayer (1L) transition metal dichalcogenides (TMDs), paving the way for on-chip polariton-based nanophotonic applications spanning quantum information processing to photochemistry.**

Polariton states, formed by the strong interaction of light and matter, embody tunable quantum properties not available in conventional material systems and have shown great promise in applications ranging from low-threshold lasing[1], efficient frequency conversion [2], and molecular sensing [3] to quantum simulation and information processing[4,5]. As polariton-supporting materials, single-layer (1L) TMDs are notable for their strongly bound exciton states (binding energies >> kT, ~200 - 800 meV[6]) that prompt large oscillator strengths and optical transitions[6,7]. These are key attributes for realizing polaritons and are tunable in 1L-TMDs using voltage, environmental changes, and strain[7,8]. However, formation of exciton-polariton states in 1L-TMDs that are controllable *in operando* has proven challenging, and has not been achieved at nanoscale sizes and under ambient conditions relevant for compact quantum-photonic devices.

Rapid progress in the integration of 2D materials and dielectric photonic elements has led to breakthrough demonstrations of strong light-matter coupling and formation of exciton-polariton states in 1L-TMDs[9-13]. In a push for miniaturization, strongly coupled plasmon-exciton polariton states, known as plexcitons, have emerged as a promising system for room temperature quantum optical applications due to ultra-confined mode volumes and corresponding fast coupling dynamics[14]. Plexcitonic systems often utilize small plasmonic cavities, termed nanocavities, which, despite significantly lower quality factors compared to their photonic counterparts, possess nanoscopic mode volumes that have allowed the observation of room-temperature plexcitons in an array of materials such as molecular J-aggregates[15], quantum dots[16,17], CdSe nanoplatelets[18] and 1L-TMDs[19].

A particularly powerful nanocavity geometry is the nanoparticle-on-mirror (NPoM), in which extremely confined gap-modes are formed between a plasmonic nanoparticle and metallic substrate[20,21]. Beyond strong electromagnetic field confinement, NPoMs offer control of the plasmonic resonances through engineering of the gap geometry and dielectric properties, enabling direct influence over both the energies and number of supported plasmonic cavity modes[22]. By combining this nanocavity flexibility with TMD excitons, the coupled NPoM/TMD system represents a promising architecture for forging and modulating plexciton states at room temperature[23,24]. However, for NPoM systems where the gap is formed by the 1L-TMD itself, native 1L excitons are nonideal for nanocavity coupling since such small gap sizes (~0.6 nm) lead to low-energy nanocavity resonances that are far detuned from common 1L-TMD exciton transitions[25]. Additional complications also arise from the predominately in-plane transition dipoles of the bright excitons[26] that are orthogonal to the dominant out-of-plane polarization of the nanocavity modes[22]. Straining the TMD can potentially overcome this difficulty. For instance, application of tensile strain can modify dipole orientations[27] redshift exciton emission energies by hundreds of meV, and tune the exciton-phonon coupling[25]. Importantly, this tuning can be dynamic and allows for *in situ* control of 1L-TMD excitons. Moreover, tensile strain can create local potential wells that trap TMD excitons into quantum dot like localized states[8,28]. These localized states can hybridize with dark excitons and host out-of-plane transition dipoles, allowing for enhanced coupling to the plasmonic nanocavity[29], making them more suitable candidates for room-temperature plexciton states.

Here we achieve *in operando* tunable strong coupling to excitons in 2D semiconductor monolayers at room-temperature enabled by a general approach combining strain engineering and adjustable plasmonic nanocavities. Using strained nanobubbles of 1L-WSe$_2$ and tip-based gap plasmons for our demonstration, we show that the exciton energy and nanocavity plasmon resonance can be controllably toggled by applying pressure with the tip, leading to observed Rabi splitting >100 meV. The results are facilitated by force spectroscopy that allows us to quantify nanocavity gap sizes and 1L-TMD indentations. Nano-PL measurements, supported by finite difference time domain (FDTD) simulations, identify distinct upper and lower polariton bands as well as uncoupled exciton or dark polariton states[30,31] and map their evolution as a function of nanogap and strain tuning. Nano-Raman spectroscopy provides an independent measure of 1L-TMD strain within the gap, showing that the observed splittings cannot be accounted for by a pure strain picture, which is further supported by strain modeling and dark polariton emission energies in the WSe$_2$. The behavior is robust under multiple cycles of applied pressure, and repeated experiments over multiple nanobubbles show that a wide array of plexciton states are accessible. Finally, we establish that the nanogap size and strong coupling can be directly modulated via an applied DC voltage between the substrate and plasmonic tip, underscoring the potential of the architecture as a plexcitonic nano-electro-mechanical system (NEMS).

**Results:**

The approach deployed in this work is based on the tools of tip-enhanced nano-optical spectroscopy, which have proven powerful for investigating 2D materials[8,32-36] (see schematics of the experimental concept in Fig. 1a,b). For our tests, 1L-WSe$_2$ is directly exfoliated onto atomically smooth, template-stripped (TS) Au[37,38], with nanobubbles forming spontaneously during exfoliation (details in supplementary materials note 1). A metalized atomic force microscopy (AFM) probe with a strong localized plasmon resonance (LPR) at its apex confines and enhances the optical field, acting as an optical antenna (Fig. 1b, left). When the metalized probe is placed over a metal substrate, the plasmon of the antenna hybridizes with its image charge, forming a coupled state in the gap between the surface of the plasmonic antenna and the mirror (Fig. 1b, right)[21,22].

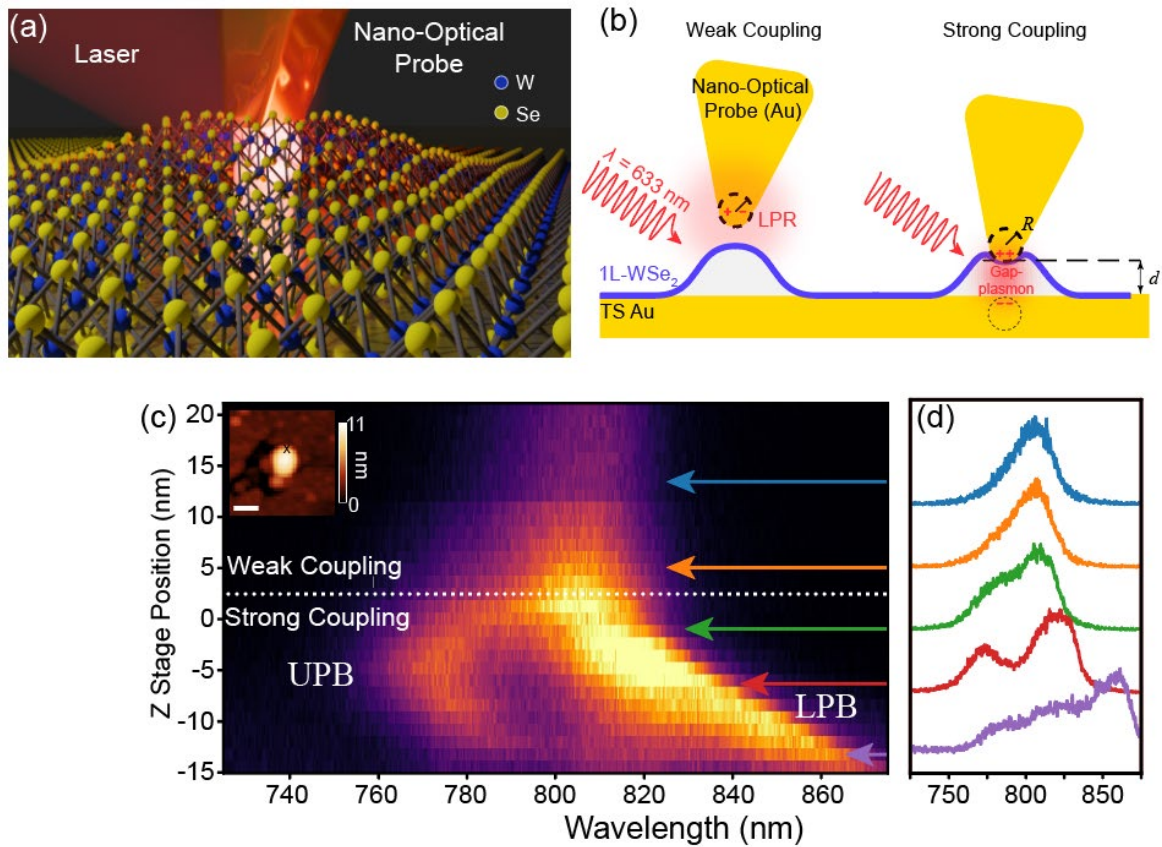

**Figure 1: Tunable strong coupling of 1L-WSe$_2$ excitons to a plasmonic nanocavity:** (a) 3D rendering of a 1L-WSe$_2$ nanobubble under indentation from a metalized nano-optical probe. A laser excites localized excitons (LXs) in the WSe$_2$, which strongly couple to a gap-mode plasmon between the probe and Au substate. (b) a cross section of the system shown in (a). When the probe is relatively far away from the nanobubble, the emission of the exciton is enhanced by the localized plasmon resonance (LPR) of the probe, but is only weakly coupled (left). When the gap size ($d$) is reduced, the gap plasmon is formed, dramatically increasing the exciton-plasmon coupling, forming a plexciton (right). (c) example of plexciton formation and tuning of LXs near an edge of a nanobubble (inset: AFM topography of nanobubble; pressing region marked by an "X"; scalebar = 50 nm). When the gap size becomes small, strong coupling is achieved, forming the upper and lower polariton branches (UPB, LPB). (d) normalized nano-PL spectra corresponding to the marked arrows in (c).

A powerful feature of these gap-mode plasmons is that mode volume $V_m$ and field enhancement are strongly dependent on gap size $d$, with $V_m \propto d^{2.5}$.[22] Therefore, as gaps approach the size of molecules, the interaction volume shrinks commensurately, accompanied by an increase in the interaction strength. The ultrasmall mode volume and large enhancements form the key to many nano-optical experiments. Additionally, it is well known from cavity quantum electrodynamics that the coupling strength of a quantum emitter and cavity field scales $1/\sqrt{V_m}$[22]. Thus, by localizing excitons inside a sufficiently small gap-mode plasmonic cavity, the coupled system can enter the strong coupling regime when the Rabi frequency is larger than the summed linewidths[19,39,40].

In this work, the emitters are localized excitons (LXs) in nanobubbles of 1L-WSe$_2$. Nanobubbles in 1L-TMDs have been extensively studied owing to their large strains and strain gradients[41], which have produced a variety of optoelectronic phenomena including exciton funneling and the activation of single-photon-emitting states at cryogenic temperatures[42]. Previous nano-optical studies have shown that the excitons in nanobubbles are strongly localized and redshifted in nanoscopic regions of high strain[8,43], with emission wavelengths 800 - 900 nm. While both nonradiative and radiative recombination rates within a sample are enhanced in close proximity to a nano-optical probe[44], a net enhancement of photoluminescence (PL), averaged over a few nm-radius from the tip-sample contact point, allows for nanoscale mapping of exciton behavior. Recent experiments have established that LXs can originate from the manifold of exciton states, including charged excitons (trions) and dark excitons, that exist with energies near the A-exciton (AX) resonance[33,36]. Following previous tip-enhanced and NPoM studies, we attribute the LXs observed here to be dark excitons[45,46].

At equilibrium, nanobubbles are under tensile strain induced by a mechanical load in the normal direction, where the load is the pressure from the trapped material between the monolayer and substrate[47]. As a consequence, the strain on the monolayer will be affected by changing the mechanical pressure (*e.g.,* by applying a force with a plasmonic nano-optical scan-probe). Since exciton transitions in 1L-WSe$_2$ are strongly affected by the strain[43], the ability to tailor the local strain provides a direct means to tune the exciton energies and coupling strengths within a nanocavity. An example of the evolution of the nano-PL spectrum over the course of the approach of a plasmonic probe towards a nanobubble and the subsequent formation of a nanocavity is shown in Fig. 1c. At first, the probe is close to the nanobubble, which itself is sitting on the Au surface, but substantial pressure is not exerted by the probe on the nanobubble. Excitons in the nanobubble are excited by the LSP fields of the probe alone (see Fig 1b, left), enhancing the photoluminescence intensity of the excitons in the nanobubble. As the tip-substrate gap is further closed by lowering the probe closer to the nanobubble, a nanocavity "gap-mode" plasmon is formed where the cavity resonance depends on the size of the gap. Eventually, the probe contacts the nanobubble and applies a mechanical pressure to the point of contact. This pressure additionally modifies the LX energy, and combined with the gap-dependent plasmon resonance, the detuning between the exciton and the nanocavity is altered. A continued reduction in gap size increases the plasmon-exciton coupling while maintaining a near-constant detuning (*vida infra*), facilitating strong coupling and forming plexciton states.

Spectroscopically, this progression can be seen in Fig. 1d. The uncoupled LX emission is shown by the blue spectrum. With weak coupling, the PL spectrum (orange) is similar, though as the gap size is reduced, a large increase in the PL intensity is observed. This increase is followed by a splitting of the PL peak into two distinct states, which we assign as the upper polariton (UPB) and lower polariton (LPB) branches (green) that signify a transition from the weak coupling to the strong coupling regimes. In these studies, the observed photoluminescence from the UPB is enabled by the large enhancement of the emission by the nanocavity. Notably, during this stage of the approach, the probe has snapped into contact with the nanobubble, but the applied pressure is net *negative* due to attractive forces of the probe-sample interaction. From an instrumentation standpoint, the probe-sample distance is controlled by a piezoelectric positioner (i.e., the z-piezo) and as the piezo-determined distance further decreases, the mechanical interaction between the probe and the surface transitions from attractive to repulsive. This transition point corresponds to the red spectrum in Fig. 1d, where an even larger splitting emerges and two new states are clearly resolved in the photoluminescence spectrum. Further reduction of the probe-sample distance from here causes the nano-optical probe to indent the sample, and a more complex spectrum with multiple emission peaks spanning from 775 to 875 nm emerges. These trends are observed on all studied nanobubbles on TS Au substrates, but do not occur for nanobubbles on purely dielectric substrates[8]: the signatures of strong coupling behavior are only observed when a nanocavity is formed.

The multifaceted spectral behavior reflects the high tunability of the nanobubble and gap plasmon system. To gain a full understanding of the system, a detailed analysis of both the gap-dependent plasmons and the nanobubble exciton resonances is needed. For instance, the known sensitivity of excitons in 1L-WSe$_2$ to strain raises the possibility that all of the observed behavior results from a complex inhomogeneous strain field, with only weak coupling to the cavity. To eliminate this possibility, independent measurements of tip-induced strain effects are needed (see below). Further, gap-mode plasmons are highly sensitive to their geometry[22,48], and thus an accurate estimation of the size of the tip-substrate gap is also critical to the analysis. Determining the size of the gap is challenging because once the tip begins interacting with the sample, the change in gap is no longer reflected by the change in the z positioner of the instrument, as the probe and sample will both deflect in response to normal forces on each.

Therefore, to estimate the gap size, we perform force spectroscopy to extract the sample and tip z-position shifts at each z-piezo value, calibrated using force-distance curves measured on a hard sapphire surface where indentation is negligible[47]. As shown for another representative nanobubble (Fig. 2), the calibration from the force spectroscopy then allows us to report the measured nano-PL spectra as a function of the size of the gap between the tip and underlying TS Au film (Fig. 2b). From the calibrated data, as the probe approaches the nanobubble, the attractive tip-sample interactions such as capillary forces first pull the cantilever toward the surface, causing changes in the gap size that are larger than the applied z-piezo movement (Fig. 2a center inset, and Fig. 2b, top section of the curve). As the approach progresses, the interaction forces go from being net attractive to repulsive, with the cross-over marked by where force values change sign from negative to positive in Fig. 2a, left panel. The position where the total force is zero is taken to occur at a tip-substrate gap corresponding to the nanobubble height, as determined separately by AFM imaging. In the repulsive regime, the probe begins to indent the nanobubble (Fig. 2a, lower inset), and the reduction in size of the gap continues, but by amounts that are smaller than the absolute z-piezo movement (Fig. 2a). This is because the cantilever also bends in response to the repulsive force exerted on it by the nanobubble and its internal pressure (Fig. 2b, bottom section of the curve). The total indentation distance into the nanobubble at a specific force is given by the difference between the measured z-piezo position when pressing on the nanobubble and the z-piezo position from the calibration approach curve on sapphire at the same force. Finally, the gap size is estimated as the indentation distance of the probe into the nanobubble subtracted from the nanobubble height (See SI Fig. 1 and supplementary note 1 for further details).

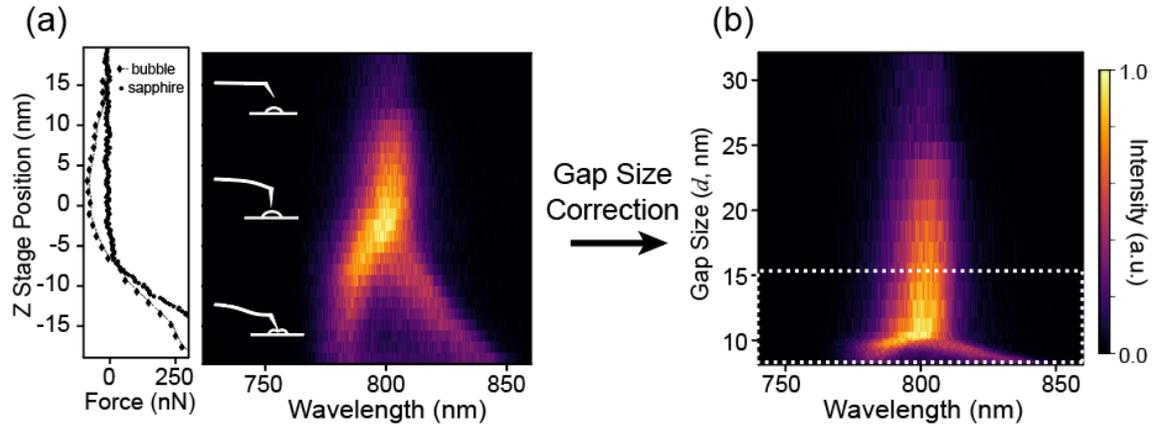

**Figure 2: Correction of z-values by estimation of gap size:** (a) nano-PL spectral evolution as a function of the raw z-stage piezo position values (right), with the corresponding Force-Distance curve (left) plotted with the calibration on a sapphire substrate. (b) the same data from (a), but now plotted versus the determined gap size, $d$. The white dotted box marks the region highlighted in Fig. 3.

Highlighted in Fig. 3 are the spectral changes that occur in Fig. 2b for the subset of gap sizes between 8 and 15 nm. As in Fig. 1c,d, the nano-PL evolution (Fig. 3a) from uncoupled nanobubble LXs (Fig. 3b, gray) to strongly coupled plexcitons (Fig. 3b, red) is observed. In addition to upper and lower polariton branches in the strong coupling regime, examination of the spectra shows remaining intensity at or near the energy of the uncoupled LX (Fig. 3a, green symbols; 3b, red curve and fits). Such a spectral feature is expected and is attributed to LXs that are not cavity-coupled or are coupled into dark polariton modes[30].

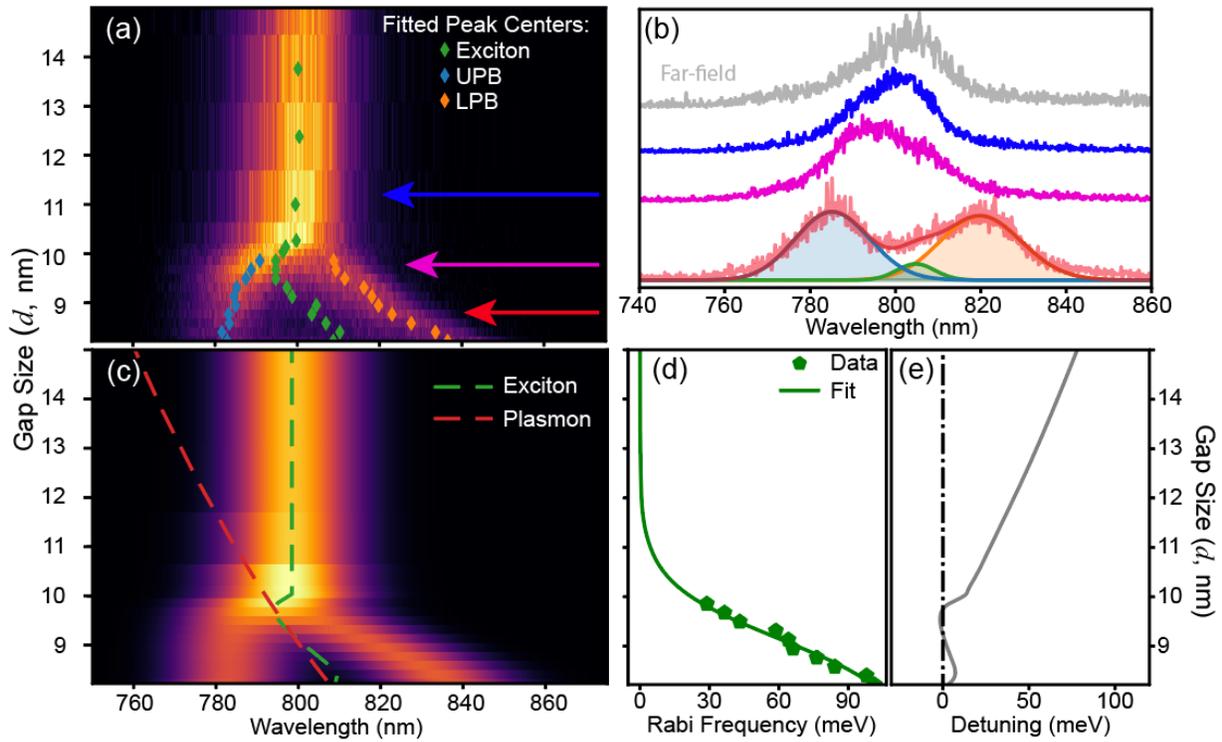

**Figure 3: Coupled oscillator modeling and Rabi frequency estimation:** (a) nano-PL spectra vs. nanocavity gap size $d$ for the subset of gap sizes marked by the white dashed box in Fig. 2b. Colored markers are the centers of 3-gaussian peak fits for each spectrum. (b) normalized individual spectra for the corresponding colored arrows in (a), showing the emergence of the UPB and LPB. Below the red curve are components of a the 3-gaussian peak fit. (c) Coupled oscillator model fitted to the data in (a) and (b). The green dashed line is the piecewise polynomial fit to the uncoupled exciton energy, and the red dashed line is the circuit model fit of the gap plasmon mode (see supplementary Fig S9 and note 3). (d) experimentally determined and model-fitted Rabi frequencies from the 3-peak fit in (a). (e) fitted detuning from the extracted plasmon and exciton energies.

Motivated by these spectral features, we use a three-peak fit to extract the positions of the UPB, LPB and uncoupled LX energies, with the fit results overlaid on Fig. 3a. The UPB and LPB are clear in the measured data and reflected by the fit. The uncoupled exciton emission initially shows a blueshift, followed by a redshift as the gap is further reduced. This is consistent with the expected strain progression from the tip-applied local force, as well as with previous near-field pressing studies[33,49]: the pressing tip initially counteracts the pressure inside the tensile-strained nanobubble, leading to a relative blue-shift of the exciton. Applying additional force eventually results in indentation, leading again to tensile strain that increases with the indentation depth. We are able to additionally verify these tip-induced strain modifications – and the strain state of the WSe$_2$ at each tip position – using tip-enhanced Raman spectroscopy (TERS) while pressing on nanobubbles (Figs. S3, S4, and Methods[41,50 51]).

Notably, the initial splitting of the PL upon the approach of the tip, which includes a strong, red-shifted LPB state, correlates with a small but statistically significant increase of the energy of the dominant Raman modes of 1L-WSe$_2$ (Fig. S3). Since compressive strains are known to cause relative Raman blueshifts[49,52], this observation allows us to rule out the possibility that the low-energy LPB state originates from a tensile-strained LXs. Instead, the relative compressive strain shifts the LX exciton energies into resonance with the plasmonic nanocavity (see Figs. 3a-b), and we thus attribute the PL splitting to exciton-nanocavity strong coupling.

With the size of the gap determined from the force spectroscopy calibration, the effect of the nanocavity can now be evaluated using established models of plasmonic resonances for gold nanostructures (SI). The distance dependence of the resonant energies and local field strengths of gap plasmons is well-studied and can be intuitively understood as the hybridization of two plasmon modes into a dipole allowed bonding mode at lower energy, and a dipole forbidden antibonding mode at higher energy[48] (Fig. S5). As the gap size is reduced, the plasmonic tip-substrate coupling strengthens, increasing the energetic redshift of the bonding plasmon farther from that of the isolated probe (Fig. 3c, red curve). We thus expect the plasmon resonance experienced by the nanobubble LXs to be approximately that of the bare probe for large gap, and then move to longer wavelengths as the gap shrinks.

To understand how the gap affects the polariton spectrum, we utilize FDTD simulations. Full details are given in supplementary note 3. Our simulations show the energy of the plasmon resonance depends on the gap size, following approximately that of a lumped circuit for an NPoM as given in equation 2 (see also[21,22]). With the plasmon modeled, we can further simulate the plexciton state by including the LX in the dielectric function of the nanobubble. Following reference[23], we model the exciton as a Lorentzian oscillator, defined over a small volume inside the bubble. The energy of the LX manifold is set to 1.55 eV, corresponding to an emission wavelength of 800 nm. This transition energy is consistent with expectations of the energies of excitons in tensile-strained nanobubbles of 1L-WSe2 in the vicinity of a metal substrate[41] and with the observed emission in Figs. 1-3. The simulations demonstrate that strong coupling between the plasmon and exciton occurs in a model nanobubble-gap mode system, showing clear Rabi splitting of the resonance peak (SI Fig. 8) in the calculated scattering spectra of the system.

To visualize the strong coupling, in Fig 3c we fit the coupled oscillator model to the polariton emission centers in Fig. 3a. For simplicity we assumed the peak profiles are two equal width gaussians. Since the bare nanocavity has negligible luminescence, to fit the expected PL we assume that the observed luminescence is transduced through the exciton fraction of the coupled state. When then take the PL intensity to be proportional to the exciton component of the model eigenvectors. For the coupling strength, we employ a logistic functional dependence on the gap size to capture both the sharp rise of the coupling field as the probe approaches and the field quenching effects that become significant at very close distances[44]. We then use this function and fit to the observed Rabi energies, using the data and fits plotted in Fig. 3c. As can be seen, the coupled oscillator model matches well to the features observed in the experimental dataset, providing further evidence of plexciton strong coupling in our system.

The Rabi splitting from the measurements and the model (Fig. 3d) highlight a key feature of this tunable plexciton system. Namely, the strong shifts of the gap-mode plasmon resonance and the exciton energies are in the same direction and similar in magnitude over a range of forces applied by the probe. This allows the strong coupling behavior to be robust over a span of gap sizes, as the detuning stays relatively constant while the coupling strength is manipulated. This is in contrast to traditional cavity polaritons where only the cavity energy may be strongly tuned [9], or to plexcitons formed from colloidal quantum dots coupled to gap mode plasmons where the exciton energy is often fixed for a given quantum dot[16,17].

We find the tunable strong coupling to be reproducible for repeated pressings on the same nanobubble (Figs.S10 and S11), though we do occasionally observe changes to nanobubble shape after repeated pressure cycles. More generally, a survey of multiple nanobubbles over different $WSe_2$/Au samples shows that the trends described above are consistent for all investigated nanobubbles, highlighting the potential of the system as a robust quantum-photonic platform. This durability is consistent with prior AFM indentation studies on nanobubbles in $MoS_2$ and graphene[47]. Quantitatively, we find nominal differences in emission peak energies, splitting energies, and intensities for different nanobubbles, and observe more complex spectral behavior in some cases (e.g., Fig. S12). We attribute these variations to the detailed structure of each nanobubble, where precise values of strain and strain-gradients will affect both local exciton properties and nanocavity mode evolution vs. applied force.

To further investigate how local nanobubble structure affects details of the strong coupling, we measure approach curves while pressing on different positions of a single, larger nanobubble (larger than the probe tip dimensions). Because of the larger size, adjacent regions of the nanobubble will experience different mechanical pressures. Also, previous work has shown that TMD nanobubble edges can possess more strongly localized excitons compared to the interior[8]. Together, these facts suggest that variations can be expected, and this is indeed observed (Fig. S13): some areas show well-defined lambda-shaped curves similar to Figs. 1-3, some regions show peaks at different energies, while others demonstrate weaker splitting under comparable applied forces.

Overall, the results of Figs. 1-3 and Figs S10 - S13 illustrate the durability and large tunability of the plexciton states via mechanical loading by the plasmonic nanoprobe. Towards device applications, more general applicability will require on-chip control of the nanocavity-TMD system. Such control can be accomplished, for example, by using electrostatic forces induced by a DC voltage to adjust the nanocavity gap, thereby establishing a prototypical NEMS platform for tunable strong coupling.

As a first step towards demonstrating the potential viability of a standalone NEMS device that exploits the tunability discovered here, we prepared another 1L-$WSe_2$ flake on a TS Au surface. In contrast to the structure above, this device prototype also includes a thin insulating spacer layer of hBN that is 3 nm thick and electronically isolates the $WSe_2$ and any nanobubbles from the Au[14] substrate. The electronic isolation allows us to apply a displacement field between the substrate and a plasmonic probe and the resulting electrostatic forces draw the probe to/from the nanobubble, playing the role of the z-piezo.

Figure 4a presents an AFM topography image of the 1L-WSe$_2$ containing nanobubbles on the hBN/Au substrate. The nanobubble of interest has a height of 12 nm along the red dashed line (top inset, Fig. 4a). Before testing the voltage response, we first show that strong coupling is also achieved in the new sample through direct mechanical pressing, as in Figs. 1–3. Fig. 4b displays the force-controlled plexciton formation and tuning behavior for the 1L-WSe$_2$ nanobubble. As expected, like Figs. 1-3, we observe a strong exciton-plasmon coupling as the gap size decreases. Notably, the presence of the hBN spacer layer between the WSe$_2$ and Au substrate results in larger far-field PL background from the 1L-WSe$_2$ exciton states present in the area surrounding the nanobubble, which are no longer fully quenched by the Au substrate. Therefore, to better visualize the plexciton formation and tunability, we subtract the far-field background from the original near-field + far-field spectra, which creates a dip in the plot near the bare A-exciton resonance. Hence, in the approach spectra in Fig. 4b, the subtracted far-field background obscures the upper polariton branch, but the lower polariton branch is clearly observed under the applied mechanical forces, confirming the strong coupling of the plasmon and exciton.

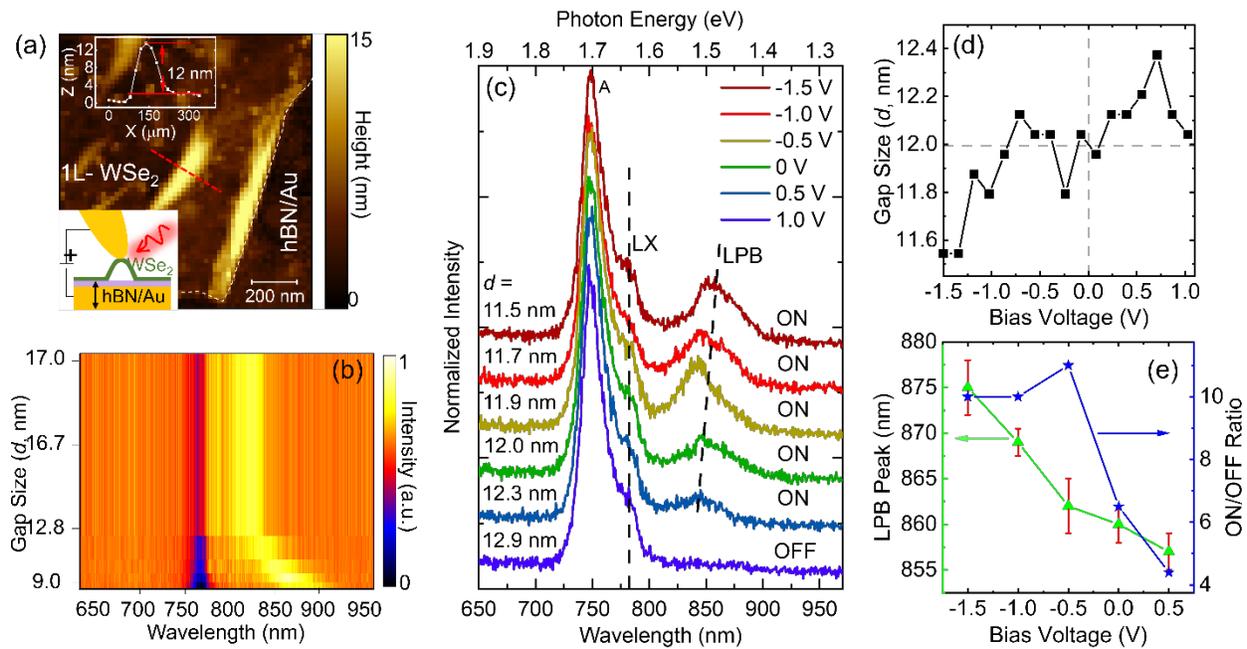

**Figure 4: Electrically controlled plexciton formation:** (a) AFM topography of extended nanobubbles in 1L-WSe$_2$ on a hBN/Au substrate, (inset top: linecut of red dashed line; bottom: device geometry of biased gap-plasmon/WSe$_2$ system). (b) Nano-PL spectra vs. gap size for mechanical pressing of nanobubble, showing a strong LPB splitting from the free exciton. (c) Nano-PL spectra vs. applied bias, showing electrical control of strong coupling. At positive biases >1 V, no LPB is observed. As the bias becomes negative, the gap size reduces, forming the plexciton. (d) corresponding gap sizes as a function of bias voltage. (e) Lower polariton branch (LPB) position and ON/OFF ratio as a function of bias voltage. We normalized all the ON spectra with respect to the OFF spectra in (c) to determine the ON/OFF ratio (e).

On the same nanobubble, we then fix the applied mechanical force with the tip in contact with the nanobubble and apply a DC bias between the probe and the substrate. The Au surface is kept at ground (see bottom inset, Fig. 4a). Figure 4c presents the corresponding nano-PL spectra at different bias voltages. Upon application of a DC bias, a static change in tip position occurs as directly reported by the tip deflection signal,[53] resulting in a tunable cavity gap. By monitoring the tip deflection under applied bias, the force spectroscopy-based calibration above enables us to determine the size of the plasmonic cavity for each bias voltage. Figure 4d shows a representative plot of the DC-bias-dependent gap size. The observed trend agrees well with results presented in Figs. 1-3 under mechanical pressing. At zero bias, the lower polariton branch is clearly observed from the initial applied force of the probe, corresponding to a gap size of ~12 nm. As the bias is tuned negative, from zero to -1.5 V, a clear LPB redshift is observed corresponding to a decrease in $d$ as measured by a change in tip deflection, mirroring the mechanical pressing data in Figure 4b. Likewise, the positive bias increases the gap size, leading to weaker coupling and blueshift of the LPB, until at 1 V only the far-field background is observed. This bias-dependent LPB tunability is plotted in Fig. 4e (left axis), where we use a Gaussian fit to determine the LPB peak center (see Fig. S14b). For this nanobubble, we find that the LPB peak wavelength can be tuned up to 20 nm via DC biasing.

In addition to plexciton spectral tunability, it is also possible to switch the plexcitonic strong coupling ON/OFF in our nanobubble systems via the electrostatic bias. Fig. 4e (right axis) plots the ON/OFF ratio of the plexciton intensity as a function of DC bias. As expected, the ON/OFF ratio increases with increasing negative DC bias and corresponding reduction in $d$. We observe a maximum ON/OFF ratio of 12 in our nanobubble system. This voltage-tunable weak-to-strong coupling behavior is repeatable (Fig. S15), confirming that the system represents a model NEMS device.

**Conclusion**

In this work, we demonstrate *in operando,* tunable plexcitonic coupling for 1L-WSe$_2$ nanobubbles within nanocavities at room temperature. Unique to the system, we show that coupling strength is fully controllable with nanomechanical movement while detuning is held nearly constant, allowing us to modulate Rabi splitting energies from 0 up to 110 meV. A suite of control measurements, modeling, and simulations elucidate the key gap-dependent nanocavity energetics and strain-modified exciton transitions involved in the coupling processes. The behaviour is robust and repeatable, and every nanobubble-nanocavity plexciton can be dynamically adjusted after fabrication. We further reveal that the plexciton coupling can be tuned electrostatically through a simple DC bias, yielding both ON/OFF device behavior and tunable absorption that can act as a foundation for optical logic gates. More generally, our results establish the viability of these states and materials platforms for room temperature, on-chip polaritonic or plexcitonic NEMS architectures.

**Methods:**

**Sample preparation:**
Template–stripped (TS) gold substrates were prepared using a cold welding technique. Using e-beam evaporation, two wafers are prepared: one with 50 nm Au and the other with 150 nm Au and 5 nm Ti adhesion layers. Wafers are then diced into 1 cm x 1 cm chips, and chips from each parent wafer are brought together in a vice, using TeX wipes to distribute the pressure and avoid chip fracture. The vice is left in a vacuum desiccator for ~30 min – 1 hour, after which the Au layers have fused together allowing the Si – template (chip without the Ti adhesion layer) to be stripped off. To avoid contamination, the template is stripped in a glove box, followed by direct scotch tape exfoliation of the WSe$_2$ onto the TS Au substrate.

**Nano optical measurements:**
Tip-enhanced photoluminescence (nano-PL) and nano-Raman measurements were conducted using the Omegascope AFM optical platform from Horiba Scientific. Nano-optical probes were gold-coated ACCESS probes from Applied Nanostructures. For nano-PL measurements, a Melles Griot HeNe 633 nm laser was used to excite the WSe$_2$ nanobubble, while for nano-Raman both 633 nm and 785 nm (Horiba Sci.) were used. Light collection was conducted using the LabRAM HR Evolution Raman spectrometer and Synapse EMCCD, also from Horiba Scientific. For optical/z-piezo measurements, incident powers 20 uW – 200 uW of 633 nm light were incident on the nano-optical probe for nano-PL measurements while Raman measurements with 785 nm used incident powers ~400 uW. Total photon integration times ranged between 100 ms – 2000 ms. Stiffness of nano-optical probes ranged form ~1 N/m to 50 N/m. Total indentation distances, however, depend sensitively on contact area and bubble internal pressure. Depending on conditions, soft probes may deeply indent a nanobubble and a stiff probe may only provide shallow indents as shown in Fig. S2.

For voltage-controlled nano-PL measurements, we applied a DC bias to the gold tip and the gold substrate was kept grounded. DC biasing and acquisition of nano-PL signal from the sample were performed simultaneously. We performed mechanical force-loading-dependent nano-PL measurements at each DC bias in which the bias was kept constant during the nano-PL recording, and vice versa. At a fixed mechanical load, the relative height change of the bubble was measured via monitoring the tip deflection (which was directly converted into a height profile in the AFM log).

**FDTD Simulations:**
Finite-difference-time-domain simulations were conducted using the Lumerical simulation software from Ansys. The simulations used uniform meshes of 0.25 nm, with perfectly matched layers boundary conditions. For the source, we employed a broadband source (400 – 1300 nm), which was used to Illuminate the nanoparticle and gap. Dielectric functions for the Au used the built-in Johnson-Cristy coefficients. The real and imaginary refractive index values for WSe$_2$ were obtained from refractiveindex.info. For the simulation of the LX state, a Lorentz oscillator was used of the form: $\varepsilon(\omega) = \varepsilon_\infty + \frac{f\omega_0^2}{\omega_0^2 - \omega^2 - i\gamma\omega}$ with $\varepsilon_\infty = 5$, $f = 0.5$, center energy $\omega_0 = 1.556 \text{ eV}$, and $\gamma = 70 \text{ meV}$, similar to those used in ref. 22, defined over a cylindrical volume (radius = 3 nm, height = 2 nm, embedded ~5nm into the nanobubble. Trapped gas was assumed to have constant index near that of water, $n = 1.33$.

**Acknowledgements:**
This work was supported by Programmable Quantum Materials, an Energy Frontier Research Center funded by the US Department of Energy (DOE), Office of Science, Basic Energy Sciences (BES), under award DE-SC0019443. K.W.C.K. acknowledges support from the DOE

NNSA Laboratory Residency Graduate Fellowship program (no. DE-NA0003960). D.J., M.R. and G.W. acknowledge partial support for this work from the Asian Office of Aerospace Research and Development (AOARD) of the Air Force Office of Scientific Research (AFOSR) FA2386-21-1-4063 and FA2386-20-1-4074 as well as Office of Naval Research award number N00014-23-1-2037. N.J.B., M.S, and E.Y. acknowledge support from the National Science Foundation through award NSF-2004437.  M.R. acknowledges support from Deutsche Forschungsgemeinschaft (DFG, German Research Foundation) for Walter Benjamin Fellowship (award no. RA 3646/1-1).

Supplementary Information for:

## *Highly tunable room-temperature plexitons in monolayer WSe$_2$ /gap-plasmon nanocavities*

Thomas P. Darlington[1,†], Mahfujur Rahaman[2,†], Kevin W.C. Kwock[3], Emanuil Yanev[1], Xuehao Wu[4], Luke N. Holtzman[5], Madisen Holbrook[1], Gwangwoo Kim[2,6], Kyung Yeol Ma[7], Hyeon Suk Shin[7], Andrey Krayev[8], Matthew Strasbourg[9], Nicholas. J. Borys[9], D. N. Basov[4], Katayun Barmak[5], James C. Hone[1], Abhay N. Pasupathy[4], Deep Jariwala[2], P. James Schuck[1*]

[1]Department of Mechanical Engineering, Columbia University, New York, NY 10027, USA

[2]Department of Electrical and Systems Engineering, University of Pennsylvania, Philadelphia PA 19104, USA

[3]Department of Electrical Engineering, Columbia University, New York, NY 10027, USA

[4]Department of Physics, Columbia University, New York, NY 10027, USA

[5]Department of Applied Physics and Applied Mathematics, Columbia University, New York, NY 10027, USA

[6]Department of Engineering Chemistry, Chungbuk National University, Chungbuk, 28644, Republic of Korea

[7]Department Chemistry, Ulsan National Institute of Science and Technology, Ulsan 44919, Republic of Korea

[8]Horiba Scientific, Novato, CA 94949, USA

[9]Department of Physics, Montana State University, Bozeman MT, 59717, USA

[†] These authors contributed equally to the work

* p.j.schuck@columbia.edu

**Supplementary Note 1: Gap Size Estimation:**

This section details how the gap size was determined from the Force Distance (FD) curves and topography of a nanobubble. The FD curve is first divided into four different sections: (1) the free section, where there is no mechanical interaction between the probe and nanobubble. (2) the attractive section, where attractive forces of the sample surface grab the tip. (3) the cross-over section, where the change in the force on the probe becomes positive, but the total force is still negative; and (4) the indentation section, in which the probe is applying a positive force on the nanobubble, where the positive force is defined in the negative z-direction. The boundary of section (3) and (4), where the total force is zero, is taken to be at the nanobubble height, and the movement of the probe is used to estimate the gap distances for each measured point in the FD curve.

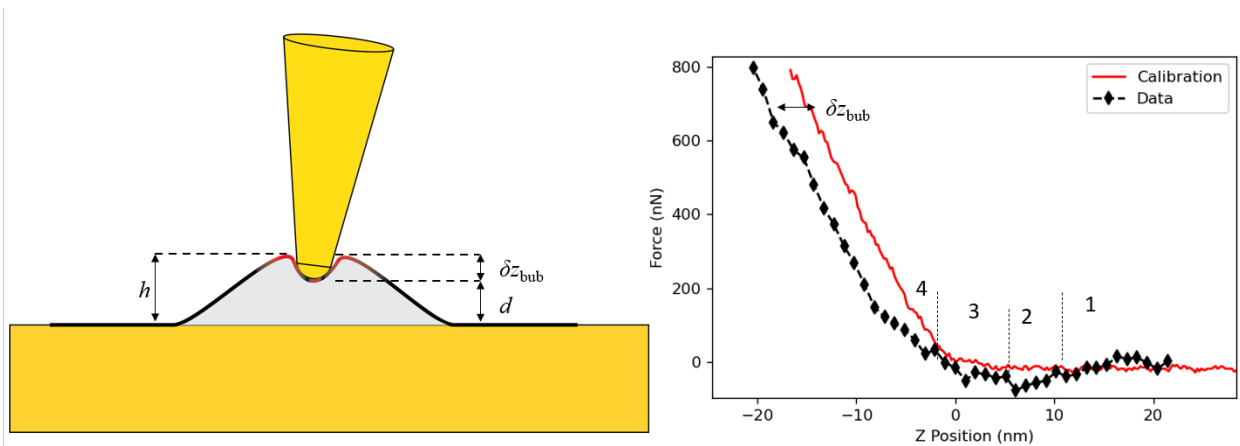

Figure S1: Gap correction procedure for FD curves presented in Figs. 2,3 of the main text. When force is positive, indentation $\delta z_{bub}$ is given by the difference between the measured $z$ position and the calibration z position at a given force. The gap is then given by difference of the nanobubble height and indentation, $d = h - \delta z_{bub}$.

In section 1 where there is no tip-sample interaction, the change in the z-piezo position should reflect the change in tip-substrate gap. In section 2, the probe begins to bend toward the sample due to attractive tip-sample interactions (the "snap-in" regime), resulting in larger deltas, until the beginning of section 3, where the minimum in the FD curve is observed. From here, the probe is nearly stationary, with comparatively small changes in the gap size, reflecting the increased force of the bubble on the probe. To the estimate the changes in the tip position in section 3, we use the approach from SI ref [1] using a Leonard-Jones model to estimate the displacement.

In section 4, the estimation of the gap size follows from measuring the indentation of the tip into the sample surface. This approach follows closely the indentation curves presented in ref 46. An FD curve for a given cantilever is first measured on a hard surface where indentation is negligible. For this experiment, a sapphire wafer was used, though standard $SiO_2$ on the Si wafers is sufficiently hard for many target samples. Because the sample is assumed to be incompressible, the z movement of the cantilever is the same as the sample, and the measured FD slope is "steep". On a soft surface however, the probe will compress the material by an amount $\delta z$. For a given force to be reached, a larger displacement of the z piezo must be applied, and the slope in FD curve is "shallow" compared to the hard sample. The amount of indentation is then estimated by aligning two FD curves and extracting the difference at a given force value. Fig S1 (left) shows a diagram of this calculation. This is then repeated for each z-stage position of the sample. With the indentation depth calculated, the gap size can be calculated from measured height of the nanobubble.

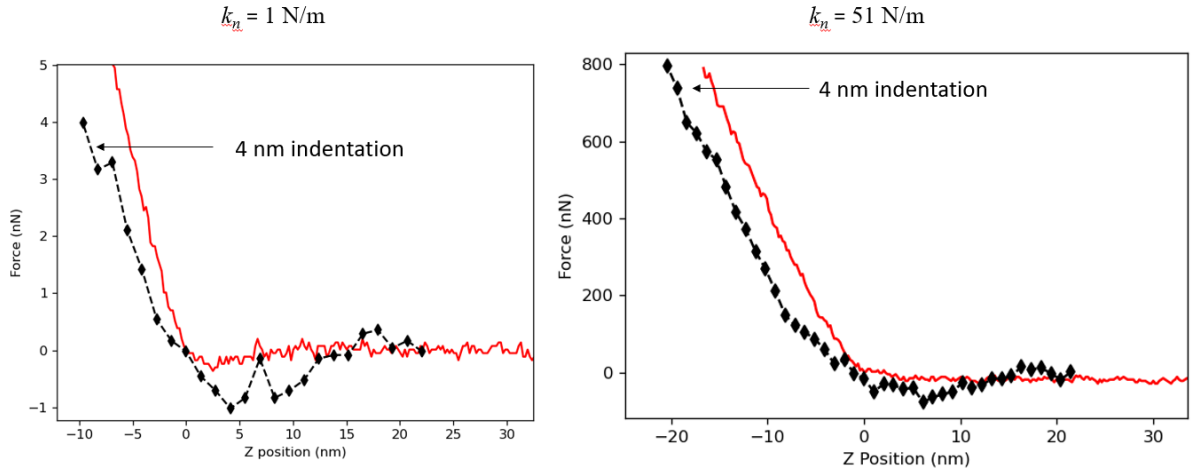

Figure S2: Comparison of indentation depths for soft and stiff nano-optical probes. Despite applying orders of magnitude lower force, the soft probe is able to ident a nanobubble 4 nm, while a much stiffer probe required orders of magnitude greater force on a different nanobubble.

**Supplementary Note 2: TERS and TEPL pressing on a single nanobubble**

Strain is well known to have a significant effect on the PL emission of $WSe_2$. As noted in the main text, it is conceivable that the spectral splitting are wholly due to strain effects. To get information on the strain state independent of the PL, we repeated the pressing on the nanobubble while observing the Raman modes. The experimental results are shown in Fig S3. The experimental approach is exactly that for the PL data shown in the main text, however the Raman signal is collected, using a high-resolution spectrometer and grating. The excitation laser used is continuous wave (CW) at 785 nm, which we showed previously as advantageous for tip enhanced studies because (a) it results in lower tip wear due to smaller the absorption of the Au tip, and (b) the sub-gap wavelength additionally reduces background from delocalized exciton emission[8]. For Raman studies of $WSe_2$, because the wavelength is below the optical gap, it also reduces the presence of 2nd order Raman peaks from the strong resonant Raman response of $WSe_2$. Fig. S3a shows the resulting Raman data as function of z-piezo stage position. Fig. S3b shows the PL for the same tip forces values with the same nano-optical probe. As can be seen, the Raman modes in Fig. S3a largely are stationary, while the nano-PL shows a clear splitting in the PL with a strong lower polariton branch. Close inspection of the Raman lines at the point of contact (blue) and large indentation (orange) shows a small (but statistically significant) stiffening of the first order Raman modes ~ 250 $cm^{-1}$, indicating compressive strain. The corresponding PL at the same probe loads are shown in Fig S3e.

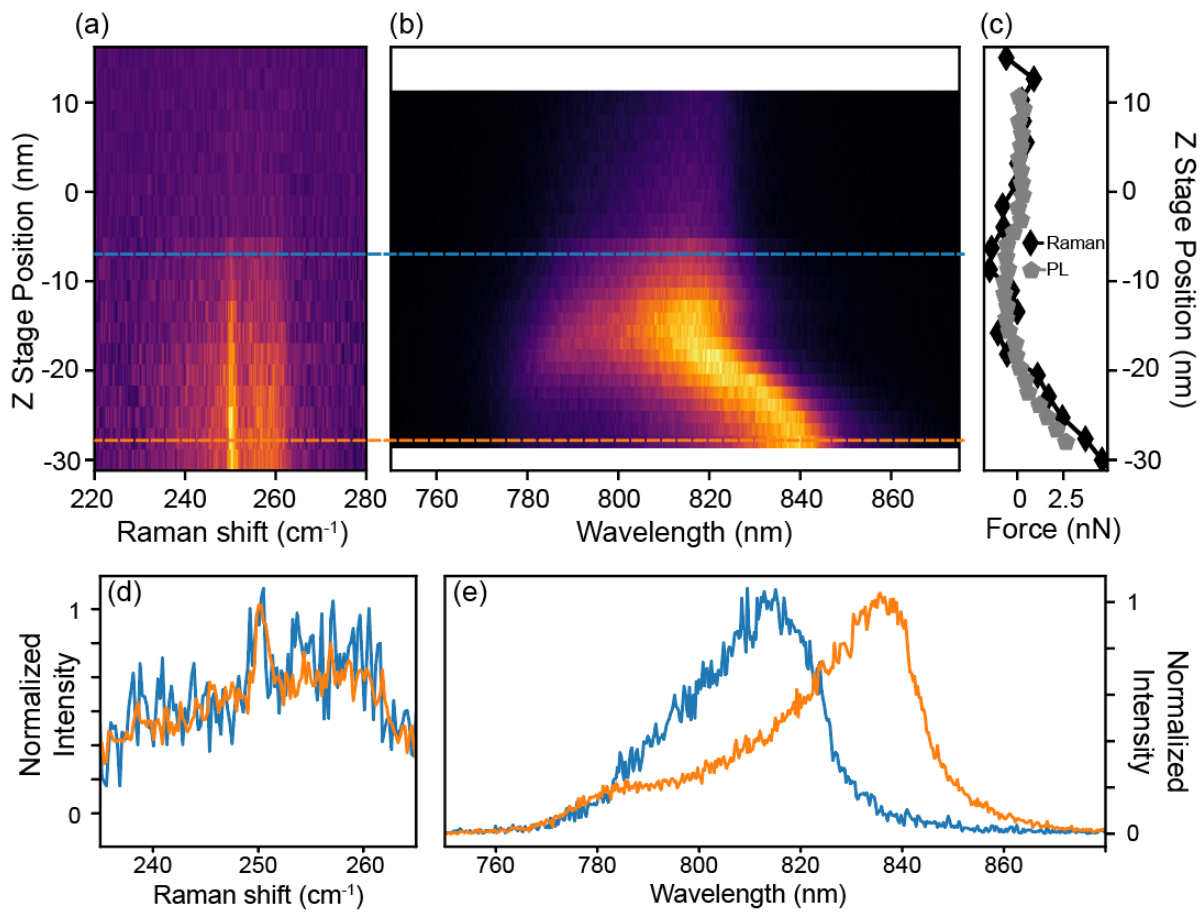

Figure S3: Nano-PL and nano-Raman approach curves on a single nanobubble. (a) Tip-enhanced Raman (TERS) using 785 nm excitation as a function of z-position approaching a nanobubble from the top. (b) Nano-PL evolution on the nanobubble. (c) The aligned FD curves for (a) and (b). (d), (e) Normalized spectra corresponding to the colored dashed lines cutting across (a) and (b). The sampe tip was used for these measurements.

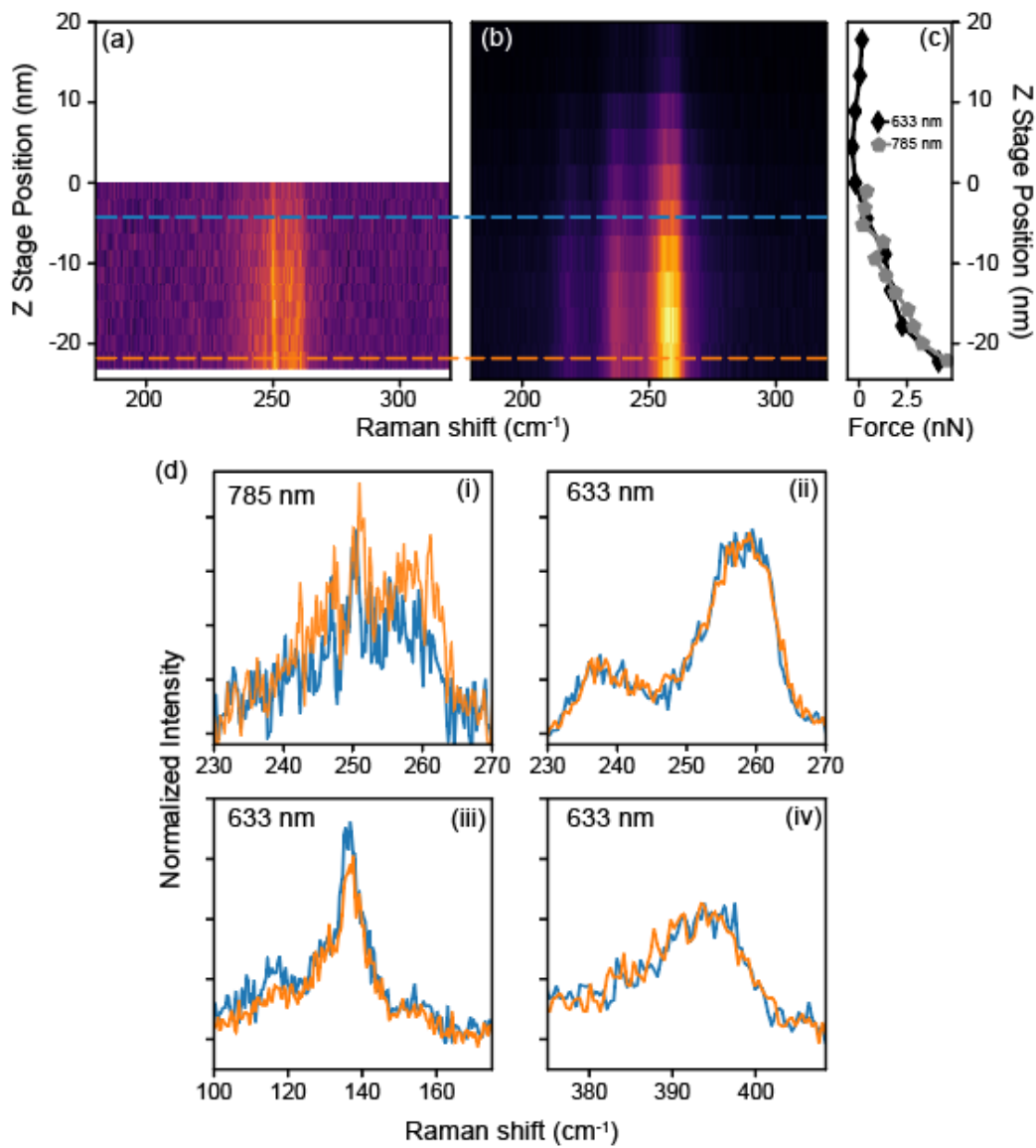

Figure S4: Comparison of nano-Raman spectra as a function of force using 785 nm (below bandgap) and 633 nm (above bandgap) excitation. (a) and (b) show the full spectra for each laser, with (c) showing the aligned FD curves for each. (d) shows normalized spectra for the dashed lines in (a) and (b) for different spectral windows. Panel (i) is centered on the 1st order Raman modes for 785 nm, and (ii) shows the 1st order Raman modes of $WSe_2$ for 633 nm. Panels (iii) and (iv) show the 2nd order Raman modes from (b).

In Fig. S4 we repeated the force-loading measurements, collecting TERS spectra using below-bandgap (Fig S4a) and above-bandgap (Fig. S4b) excitation while pressing on the same nanobubble using the same probe. As can be seen, the below-gap excitation has significantly reduced scattering from the second order Raman modes 240 and 260 cm$^{-1}$. Both Raman curves are largely unchanged with the probe forces when loaded to similar force values as Fig. S3.

The results in Figs S3 and S4 show the apparent strain is relatively small and compressive for these reported applied pressures. Since tensile strains are associate with PL redshifts, this strongly implies that the shifts observed in Fig. S3b are due to the nanocavity.

**Supplementary Note 4: Gap Plasmon Distance Dependance and Coupled Oscillator Model Fit:**

To model the plasmonic nanocavity, we employed a lumped circuit model following ref. 21 of the main text. The probe plasmon and its image are modeled as capacitively coupled LC resonators, where the value of the capacitance is a function of the gap geometry. Assuming Au for the tip plasmon and mirror and a Drude model for the metallic dielectric function, the resonance wavelength of the bonding plasmon can be approximated as:

$$\lambda_\mathrm{D} = \lambda_\mathrm{p} \sqrt{\varepsilon_\infty + 2\varepsilon_\mathrm{d} + 4\varepsilon_\mathrm{d} C_\mathrm{g}/C_\mathrm{NP}} \tag{1}$$

Where $C_\mathrm{g}, C_\mathrm{NP}$ are the capacitances of the gap and a nanosphere respectively, and $\varepsilon_\mathrm{d}$ is the dielectric function for the embedding medium, and $\varepsilon_\infty$ and $\lambda_\mathrm{p}$ are the Drude parameters for Au. Furthermore, modeling the nano-optical probe as a sphere, the value of $C_\mathrm{g}$ and $C_\mathrm{NP}$ the bonding plasmon takes on the form

$$\lambda_\mathrm{D} = \lambda_\mathrm{p} \sqrt{\varepsilon_\infty + 2\varepsilon_\mathrm{d} + 4\varepsilon_\mathrm{d} \varepsilon_g^\chi \ln\left[1 + \varsigma R/d\right]} \tag{2}$$

Where $R$ is the nanosphere radius, $\varsigma$ and $\chi$ are empirically determined numerical constants.

From the functional form of equation 2, it is easily seen that as the gap size *d* shrinks, the value of the resonance will shift to longer wavelengths, as expected from the bonding plasmon model. However, since the nano-optical tip is not truly spherical, it is not a priori obvious what value to choose for the radius, which sets resonance wavelength. A further complication is that the effective gap index is not constant with the gap, but will change as the relative volume of the air gap and the nanobubble.

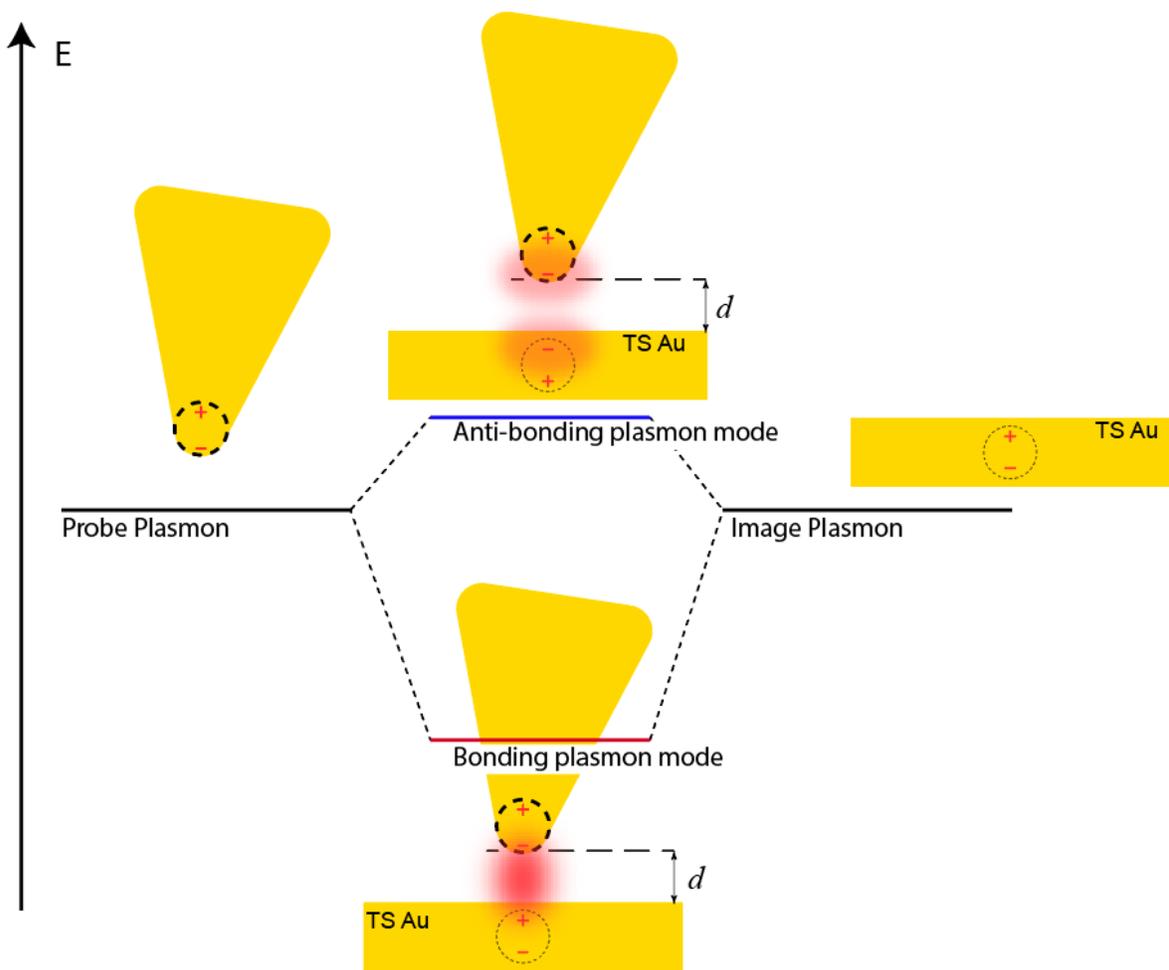

Figure S5: Illustration of gap-mode plasmon formation in the nanocavity.

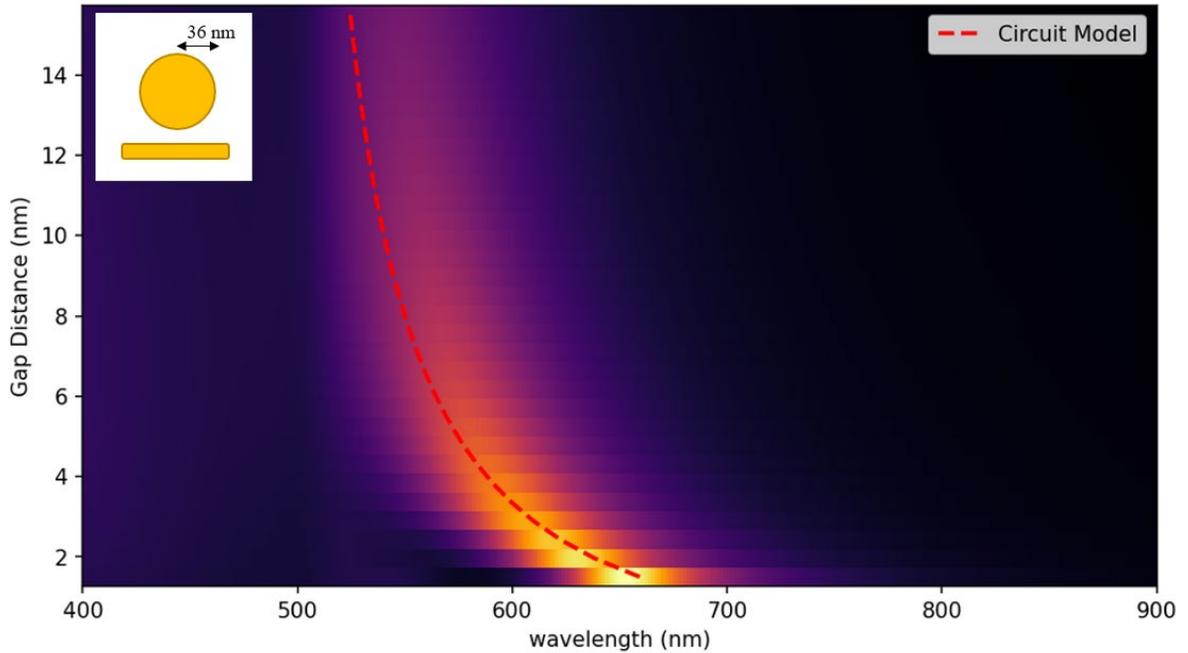

Figure S6: FDTD simulations of the plasmon response for a Au nanosphere over Au substrate with an air gap as a function gap distance. Red curve shows the lumped circuit model of equation 2.

To get more insight into the effects of these parameters on the gap dependence of the resonance, we performed full wave electromagnetic simulations of various plasmonic antennas using finite difference time domain (FDTD) methods. As a check, we first simulate a Au nanosphere in air. The scattering cross section as a function of gap size is shown in Fig S6, overlaid with the resonance wavelengths calculated from equations. As can be seen, equation 2 well describes the gap dependence, showing a nearly constant resonance until $d$ becomes small compared to the radius of the sphere. We next investigate how a modified geometry affects the gap dependence, switching the nanosphere for a nano ellipsoid of the same volume as the nanosphere. The gap dependence of the scattering cross section is shown in Fig S7a. Fitting a simulated scattering spectrum to equation 2, we see that the nano-ellipsoid scattering center qualitatively reflects that of a sphere with a larger effective radius $R = 268$ nm but with a shallower z dependence. Examining equation 2, one sees that this is equivalent to a modified $\varsigma$ pre-factor, reflecting the changed capacitance of the nanogap.

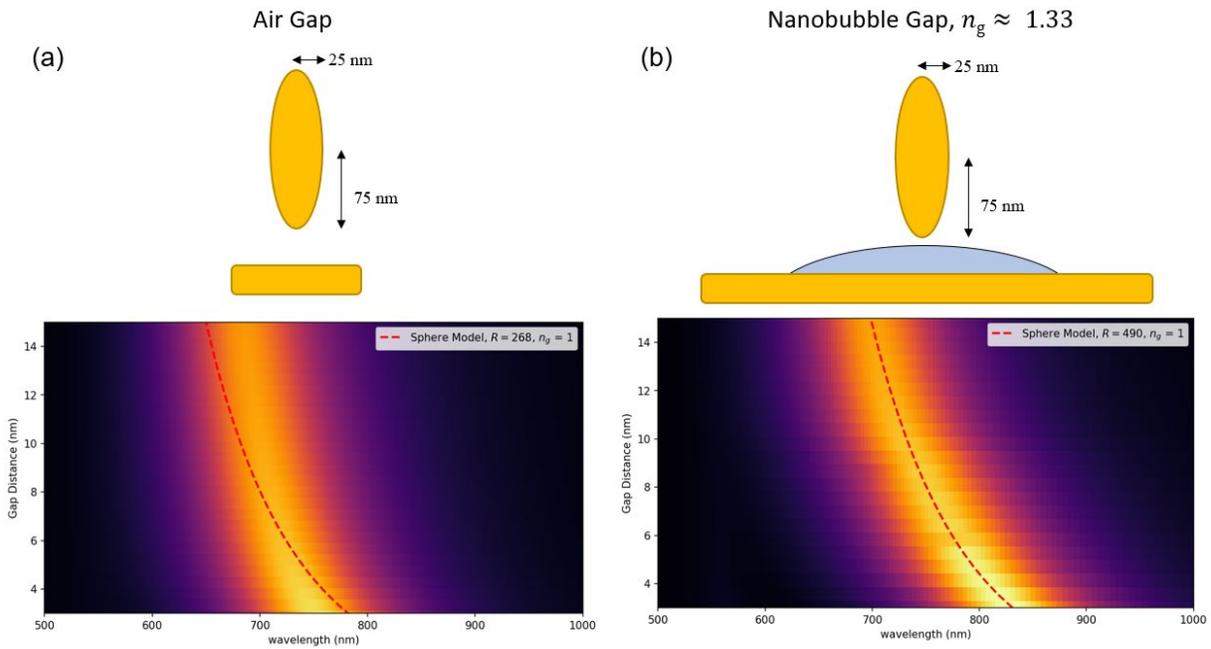

Figure S7: FDTD simulations of nano-ellipsoid/substrate gap dependence for an air gap (a) and nanobubble gap (b) compared with the lumped circuit model of a nanosphere.

To examine the changes in the index, we next place a nanobubble inside that gap. The dimensions of the model nanobubble are chosen to reflect the typical size of a curricular symmetric nanobubble as reported in main text ref. 21. An approximate index of 3 was used for the WSe$_2$ monolayer, with a thickness of 1 nm, and a value of $n = 1.33$ for the trapped material, corresponding to an average gap index of approximately 1.4 over the nanobubble. Fig. S7b shows the corresponding scattering spectra as a function of gap size. Interestingly, the nanobubble case shows even better agreement with the nanosphere case, if the gap index is allowed to vary as well as the radius. Taken together, the simulation results show that the simple equivalent circuit model can well describe the plasmon resonance shift with gap distance when used with appropriate values for the effective radius and gap index.

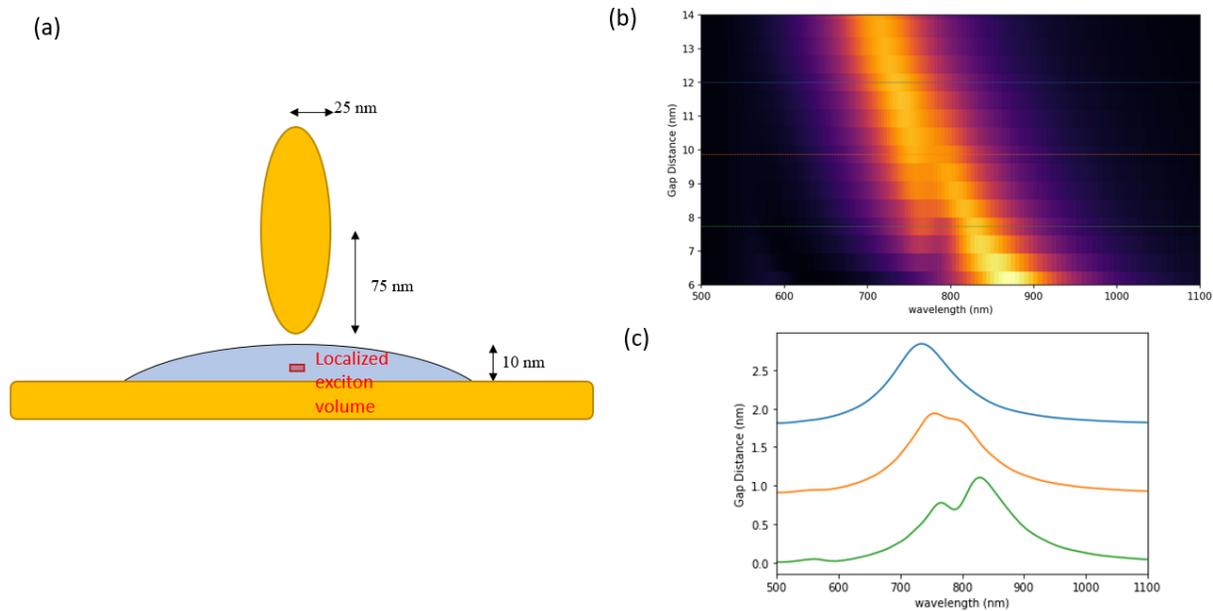

Figure S8: FDTD simulations of strong couple of a LX embedded in a nanocavity. (a) Diagram of the nano-ellipsoid over Au substrate. (b) Scattering spectra as a function of gap distance. (c) Selected spectra from the colored lines in (b).

With the gap size determined from the force spectroscopy calibration, the effect of the nanocavity can now be evaluated using established models of plasmonic resonances for gold nanostructures. As described above, the distance dependence of the gap plasmons can be intuitively understood as the hybridization of two plasmon modes into a dipole-allowed bonding mode at lower energy, and a dipole forbidden antibonding mode at higher energy (Fig. S6). As the gap size is reduced, the plasmonic probe-substrate coupling strengthens, increasing the energetic redshift of the bonding plasmon farther from the original probe resonance energy (see red curve, Fig. 3b). We thus expect the plasmon resonance experienced by the nanobubble LXs to be approximately that of the bare probe for large gap, and then move to longer wavelengths as the gap shrinks.

Our FDTD simulations show that the equivalent circuit model can approximate the gap dependence, with the value of the sphere be chosen to match the experimental plasmon resonance of the nano-optical probe. This can be done using the observed Rabi splitting in Figure 2. We assume a simple coupled oscillator model of an exciton coupled to a plasmon. With the experimental values for the upper and lower polaritons, and the uncoupled exciton energies, the plasmon energy is then

$$E_\mathrm{p} = E_\mathrm{UPB} + E_\mathrm{LPB} - E_\mathrm{Ex}$$

Which we then fit to equation 2. The fit and the result are shown in Fig. S9.

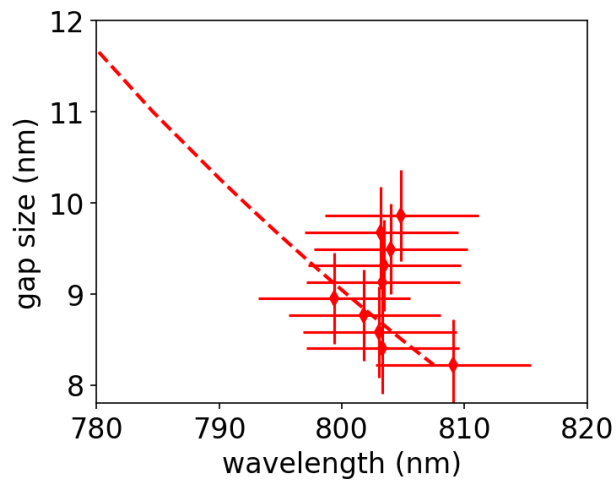

Figure S9: Fit of lumped circuit model (equation 2) to the experimental plasmon energies extracted from the fits to the upper polariton, exciton, and lower polariton in Fig. 3 of the main text.

With fitted functions for the plasmon and exciton energies, as well as the Rabi frequency, we simulate the expected luminescence of the coupled plexciton state. Assuming that the luminescence is through the exciton channel, the probability of emission, and thus the intensity, should be related to the eigenvector's exciton component. The simulated luminescence is then plotted by constructing a set of spectra with gaussians of equal width, with centers given by the two eigenvalues and heights given by the exciton component of the eigenvector.

**Supplementary Note 3: Repeated Pressings**

Fig. S10 shows repeated loading of the same nanobubble shown in Fig. 2. Two previous pressings are displayed in columns i and ii, with the corresponding Force - Distance curves plotted below. Each dataset shows remarkably similar spectral characteristics.

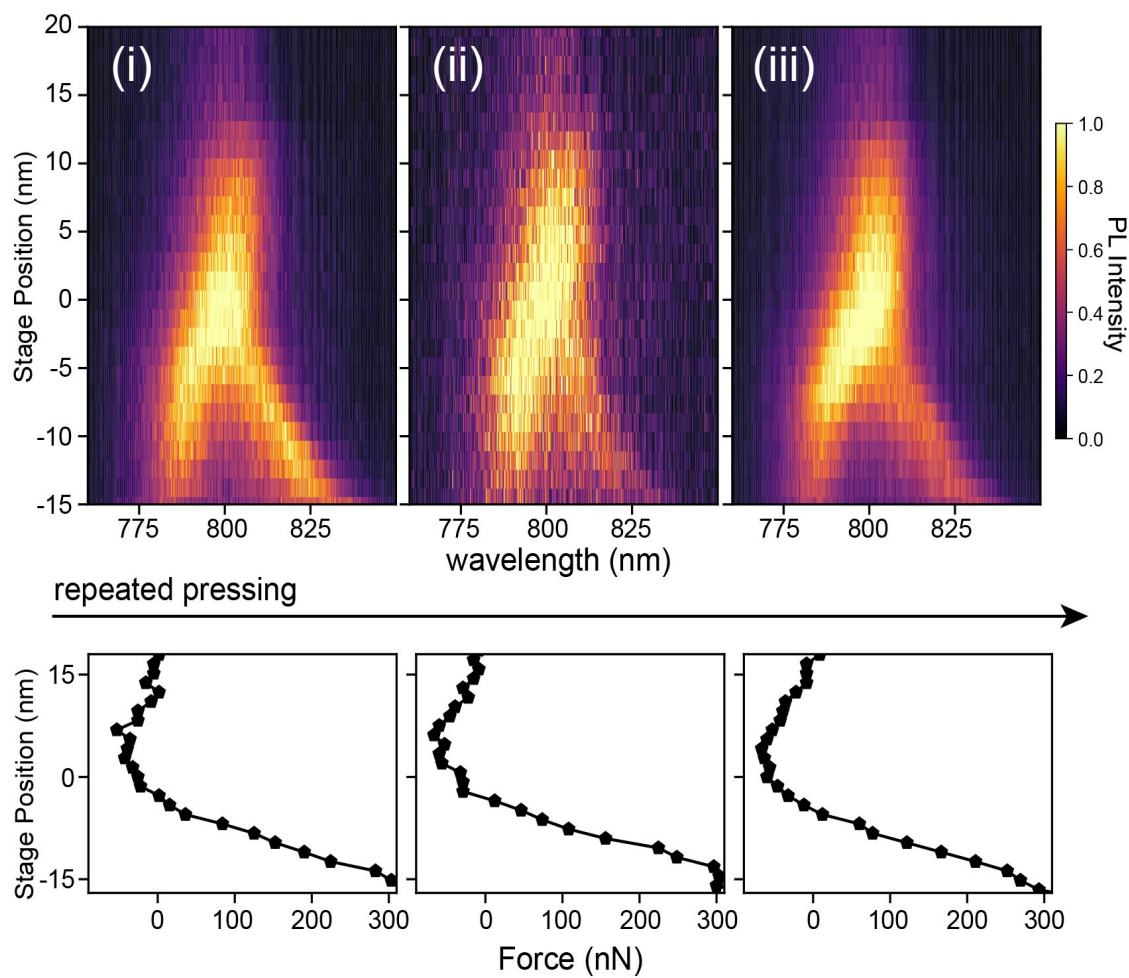

Figure S10: Repeated applied force experiments for the nanobubble presented in Figs. 2 and 3 of the main text. Top line: spectra as a function z-stage position. Bottom line: corresponding FD curves. Excitation laser powers from left to right were 200 uW, 20 uW, and 200 uW.

To further display this reproducibility, in Fig. S11 we show additional repeated force-loading PL curves on 3 different $WSe_2$ nanobubbles. In Figs. S11a and S11b, each bubble shows very similar spectral splittings, even under complete indentation, which is marked by the sudden spectral shift to a constant value coinciding with increased TERS signal. In Fig. S11c, we show a single nanobubble subject to 6 repeated loadings, with the maximum load increased each cycle. Each iteration shows remarkably similar spectral changes.

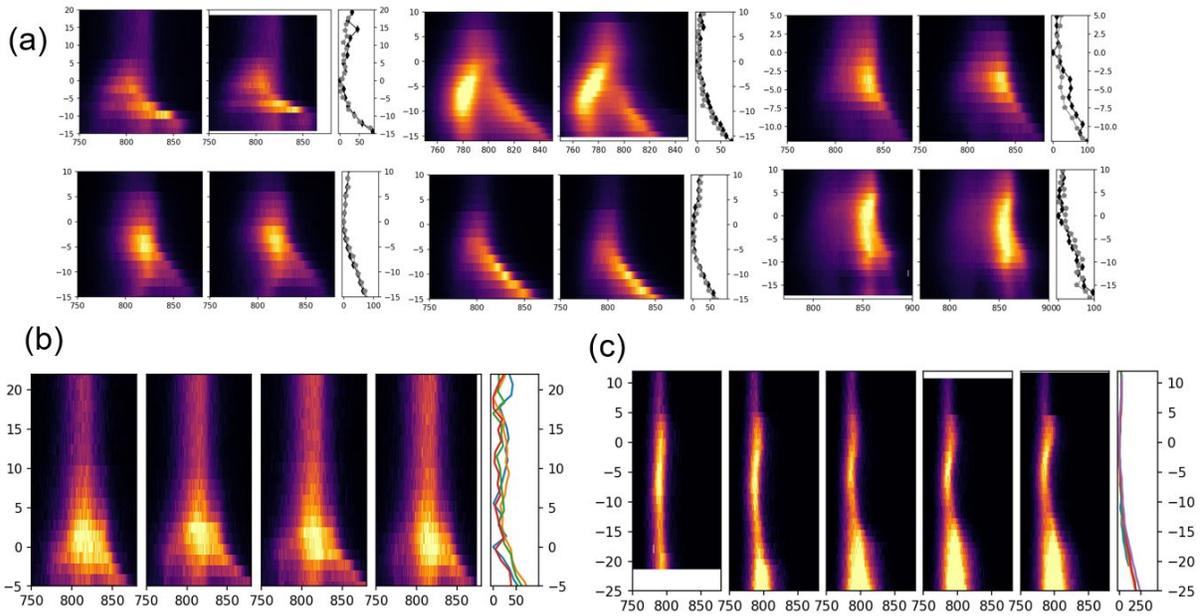

Figure S11: Repeating loadings on additional nanobubbles. (a) pair loadings for 6 nanobubbles. (b) and (c) 4x and 5x repeating loadings for nanobubbles. The x, y axes of all figures are in units of nanometers.

While spectra in Fig. S11c show changes with the applied force, no clear peak splitting is seen. Examinations of the force-distance curve show that the indentation reaches 4 nm and stays steady, indicating that the induced strain of the $WSe_2$ nanobubble likewise ceases to increase with more force. This further implies that the gap size and coupling remain constant as well. However, we note that the peak shows significant broadening as the force exceeds 500 nN, which suggests that greater indentation would reveal the upper and lower polariton branches in the PL.

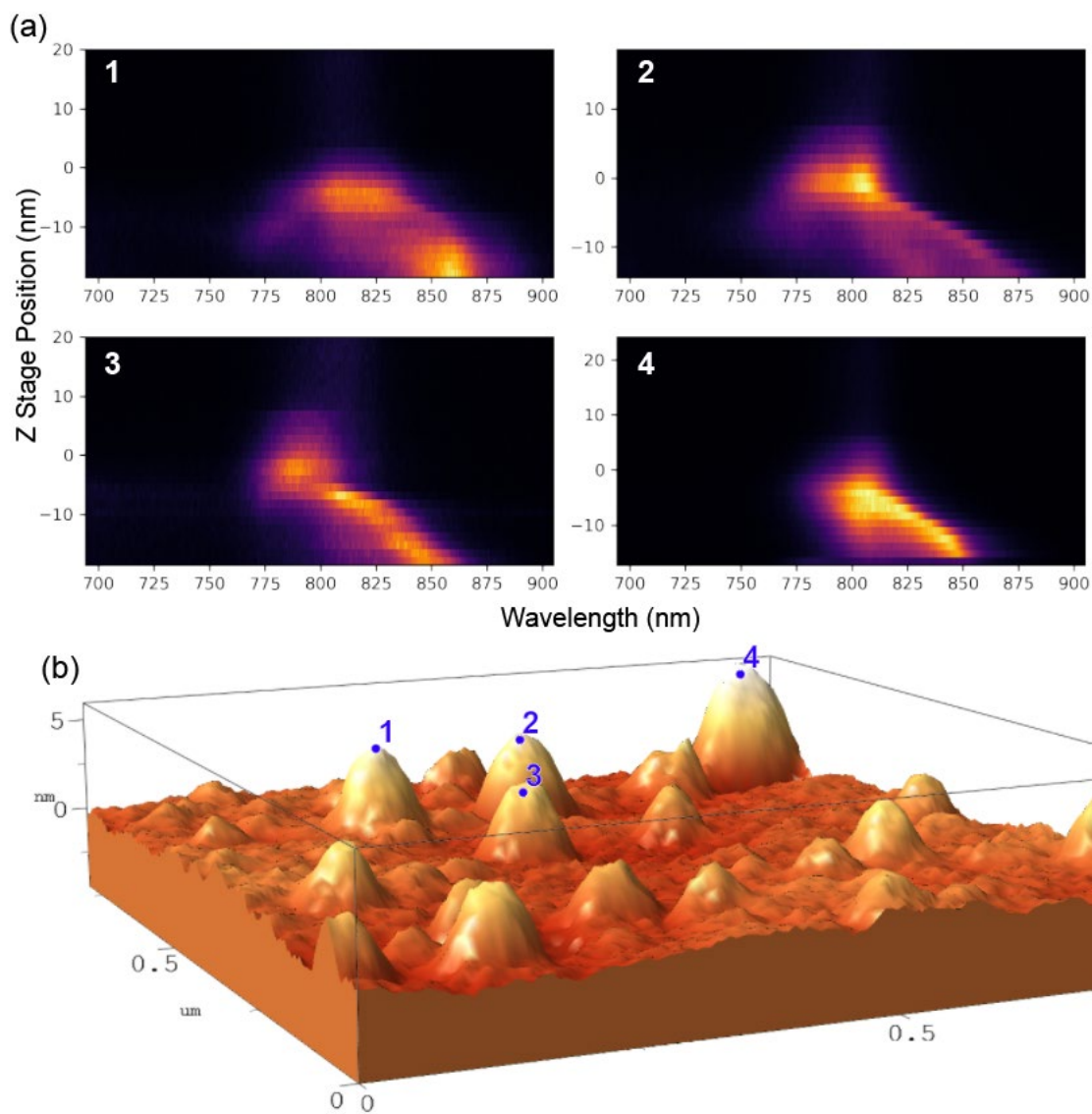

Figure S12: Force loadings on different nanobubbles in a field. (a) Nano-PL spectra as a function of z-stage potion. (b) AFM topography of the field, with each studied nanobubble labeled.

We investigated the effect of the applied mechanical force loading on several small nanobubbles of similar size. Fig. S12 shows the atomic force topography of the nanobubble field, recorded by the nano-optical probe. Fig. S12a show the corresponding nano-PL distance curves. From the AFM, each nanobubble shows roughly similar morphology as measured by the plasmonic nano-optical probe. The similar size and profile suggest that the response to the applied load of each nanobubble should likewise be similar, which should be reflected in the nano-PL emission as the localized excitations become strongly coupled. This is indeed the case in the nano-PL vs. distance curves, which show qualitatively the same behavior; the UPB and LPB emerging shortly after the probe comes into contact with the nanobubble, and a strong redshift of the LPB when the gap is further reduced. However, there are notable differences in the fine structure: each nanobubble for instance shows slightly different LX energies in the weak coupling region, likely reflecting differences in the native nanobubble strain before the tip applies any force.

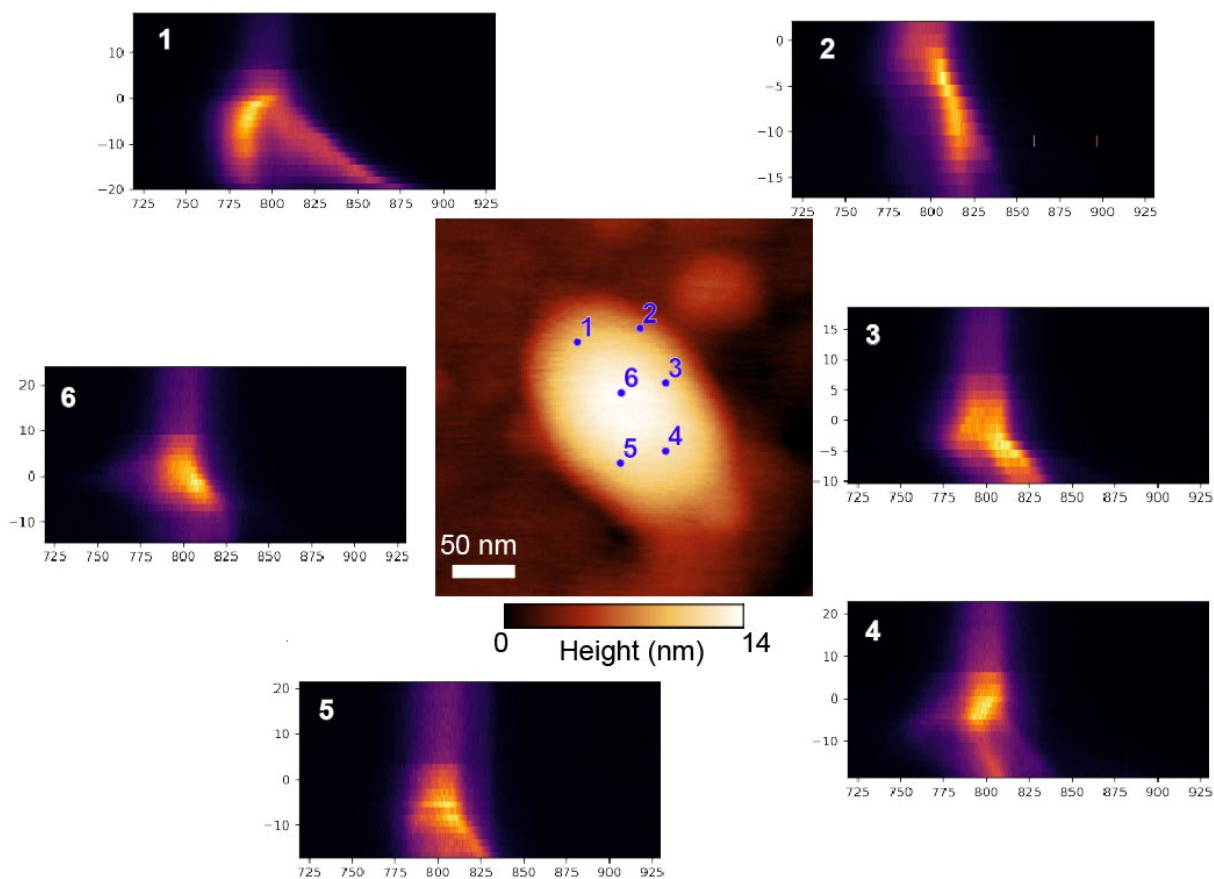

Figure S13: Plexciton formation in different areas of a single nanobubble.

**Supplementary note 4: Electrostatic control of plexciton tuning**

Fig. S14a presents the DC-bias-dependent relative height change of the WSe$_2$ nanobubble on hBN/Au substrate (i.e., the relative change in tip-substrate gap size). At 0 V bias, deflection of the tip cantilever determines the relative cavity height at 1.3 nm. Fig. 14a presents height profile for DC bias sweep from negative to positive direction. We also measure the height profile for positive to negative sweep and observe similar results. As discussed in the main text, positive (negative) bias increases (decreases) the cavity gap size. We use a linear interpolation of the plot to determine the cavity gap size shown in Fig. 4c.

Fig. 14b presents representative deconvolution of two TEPL spectra acquired at -1.5 V and 0 V, respectively. We use 4 Gaussian peaks to deconvolute the spectra. The second LPB peak is plotted as a function of the applied DC bias in Fig. 4e.

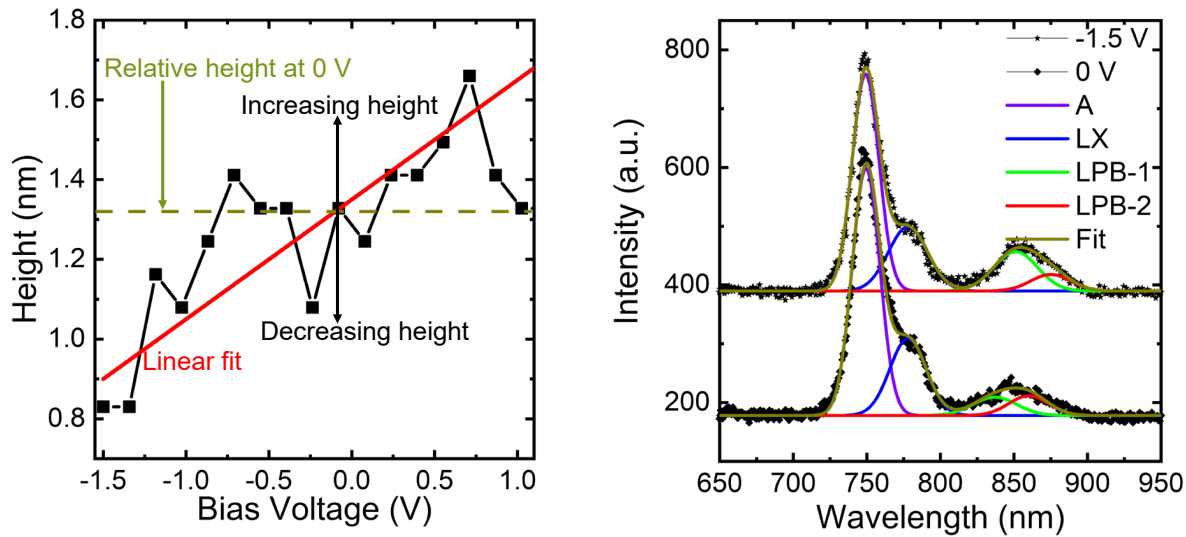

Figure S14: Electrostatic tuning of nanocavity. (Left) Relative cavity height (i.e., gap size) change as a function of bias. (Right) Spectra a 0 V and -1.5 V with corresponding excitonic contributions.

**Supplementary note 5: Reproducibility: Electrostatic control of plexciton tuning**

Fig. S15 presents three repetitive TEPL approach graphs under electrostatic bias. We start the measurement by acquiring approach spectra on the nanobubble at 0 V followed by at a positive bias (1 V) and then repeating the experiments at 0 V again. As shown, we observe plexciton switching and tunability at both 0 V bias and no trace of plexciton formation at 1 V. This demonstrates the reproducibility and robustness of plexciton manipulation and suitability of $WSe_2$ nanobubbles as plexcitonic NEMS device components.

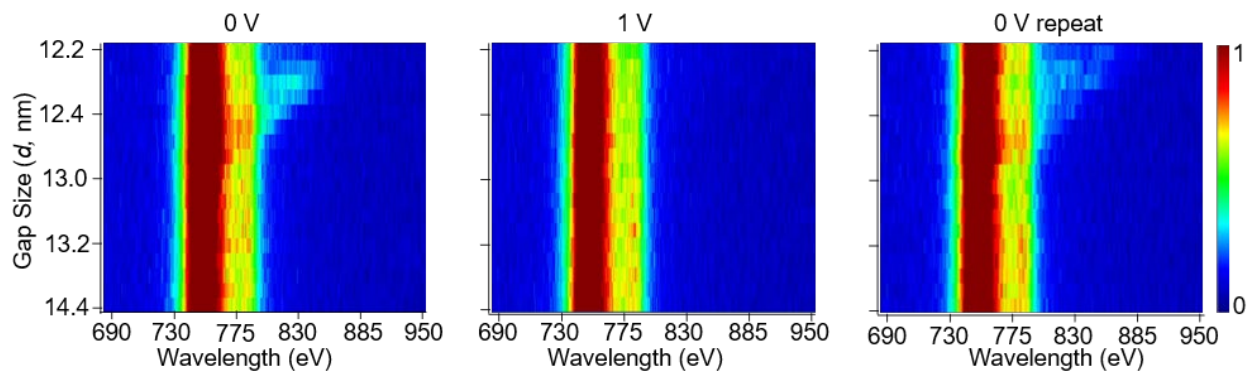

Figure S15: Reproducibility of electrostatic tuning.